\documentclass[preprint,aps,floats,nofootinbib]{revtex4}

\input epsf.tex
\def\DESepsf(#1 width #2){\epsfxsize=#2 \epsfbox{#1}}

\usepackage{graphicx}

\def\l {\lambda}

\def\msnus {m_{\tilde\nu_{iL}}^2}
\def\msells {m_{\tilde e_{iL}}^2}

\newcommand{\no}{\nonumber\\}

\newcommand{\be}{\begin{equation}}
\newcommand{\ee}{\end{equation}}
\newcommand{\bea}{\begin{eqnarray}}
\newcommand{\eea}{\end{eqnarray}}

\newcommand{\pr}{\prime}

\def\thebibliography#1{\centerline{\bf REFERENCES}
  \list{[\arabic{enumi}]}{\settowidth\labelwidth{[#1]}\leftmargin
  \labelwidth\advance\leftmargin\labelsep\usecounter{enumi}}
\def\newblock{\hskip .11em plus .33em minus -.07em}\sloppy
  \clubpenalty4000\widowpenalty4000\sfcode`\.=1000\relax}

\begin{document}
\preprint{\vbox{\hbox{}\hbox{}\hbox{MIFP-05-22}}} \draft

\vspace*{0.5cm}

\title{The $B \to K\pi$ Puzzle and Supersymmetric Models}

\author{ \vspace{0.5cm}
Richard Arnowitt$^1$\footnote{arnowitt@physics.tamu.edu},
~Bhaskar~Dutta$^1$\footnote{dutta@physics.tamu.edu}, ~Bo
Hu$^2$\footnote{bohu@ncu.edu.cn} ~ and ~
Sechul~Oh$^3$\footnote{scoh@phya.yonsei.ac.kr}}

\affiliation{ \vspace{0.3cm}
$^1$Department of Physics, Texas A\&M University, College Station, TX, 77845, USA \\
$^2$Department of Physics, Nanchang University, Jiangxi 330047, China \\
$^3$Natural Science Research Institute, Yonsei University, Seoul
120-479, Korea} \vspace*{0.5cm}

\begin{abstract}
\noindent
In the light of new experimental results on $B \to K\pi$
decays, we study the decay processes $B \to K \pi$ in the framework
of both R-parity conserving (SUGRA) and R-parity violating
supersymmetric models. We find that any possible deviations from the
Standard Model indicated by the current data for the branching
ratios and the direct CP asymmetries of $B \to K\pi$ can be
explained in both R-parity conserving SUGRA and R-parity violating
SUSY models. However, there is a difference between the predictions
of both models to the time-dependent CP asymmetry observable
$S_{K_{_S} \pi^0}$ whose current experimental results include large
uncertainties.  We demonstrate that this difference can be useful
for testing both models with more accurate data for $S_{K_{S}
\pi^0}$ and $A_{CP}^{+-}$ in the near future.
\end{abstract}
\maketitle


The quark level subprocesses for $B \to K\pi$ decays are $b \to s q
\bar q ~ (q = u,d)$ penguin processes which are potentially
sensitive to any new physics effects beyond the Standard Model (SM).
All the $B \to K\pi$ modes have already been observed in experiment
and their CP-averaged branching ratios (BRs) have been measured
within a few percent errors by the BaBar and Belle collaborations
\cite{HFAG,Bornheim:2003bv,Chao:2003ue,Aubert:2002jb,Aubert:2004dn,Aubert:2004km,Aubert:2004kn}.
The measurements of CP asymmetry observables for the $B \to K\pi$
modes had contained large errors so that the results have not led to
any decisive conclusions until recently
\cite{HFAG,Abe:LP05,Chen:2000hv,Aubert:2004qm,Chao:2004jy,Chao:2004mn,Abe:2004xp}.
But, the direct CP asymmetry in $B^0 \to K^{\pm} \pi^{\mp}$ has been
recently observed at the 5.7$\sigma$ level by BaBar and
Belle \cite{Aubert:2004qm,Chao:2004jy,Chao:2004mn} whose values are
in good agreement with each other: the world average value is
\begin{equation}
{\cal A}_{CP}^{+-} = -0.119 \pm 0.019 ~.
\end{equation}
The direct CP asymmetry data for the other $B \to K\pi$ modes still involve large
uncertainties: e.g., for $B^{\pm} \to K^{\pm} \pi^0$ modes,
${\cal A}_{CP}^{+0} = +0.04 \pm 0.04$.

The recent experimental data for the CP-averaged BRs of $B \to K\pi$ may indicate a possible
deviation from the prediction of the SM:
\begin{eqnarray}
R_c \equiv \frac{2\bar{\cal B}^{+0}}{\bar{\cal B}^{0+}}
 = 1.00 \pm 0.09 ~, ~~~~
R_n \equiv \frac{\bar{\cal B}^{+-}}{2\bar{\cal B}^{00}}
 = 0.79 \pm 0.08 ~,
\label{RcRn}
\end{eqnarray}
where $\bar {\cal B}^{ij}$ denote the CP-averaged BRs of $B \to K^i \pi^j$ decays.
It has been also claimed that within the SM, $R_c \approx R_n$ \cite{Buras:2004th,Buras:2003dj}.
The above experimental data show the pattern $R_c > R_n$ \cite{Buras:2004th,Buras:2003dj},
which would indicate the enhancement of the electroweak (EW) penguin and/or the color-suppressed
tree contributions \cite{Kim:2005jp}.

On the other hand, in the conventional prediction of the SM,
${\cal A}_{CP}^{+0}$ is expected to be almost the same as ${\cal A}_{CP}^{+-}$:
in particular, they would have the {\it same} sign.
However, the current data show that ${\cal A}_{CP}^{+0}$ differs
by 3.5$\sigma$ from ${\cal A}_{CP}^{+-}$. This is a very
interesting observation with the new measurements of ${\cal
A}_{CP}^{+-}$ by BaBar and Belle, even though the measurements of
${\cal A}_{CP}^{+0}$ still include sizable errors. This possible
discrepancy from the SM prediction, together with the above one on
$R_c$ and $R_n$, has recently been called the ``$B \to K\pi$
puzzle''. One may need to explain on the theoretical basis how
this feature can happen.

In the light of those new data, {\it including the direct CP asymmetry in
$B^0 \to K^{\pm} \pi^{\mp}$}, many works have been recently done to study the implications
of the data \cite{Buras:2004th,Kim:2005jp,Mishima:2004um,Charng:2004ed,He:2004ck,Wu:2004xx,
Baek:2004rp,Carruthers:2004gj,Nandi:2004dx,Morozumi:2004ea,Khalil:2005qg,Li:2005kt,Kim:2005pe}.
However, most of those previous works have focused on finding out the $B \to K\pi$ puzzle
itself and clarifying its implications through model-independent approaches, such as
the topological quark diagram approach.

In this letter, we focus on how to resolve the $B \to K\pi$ puzzle
with well-motivated new physics models: in the framework of R-parity
conserving and R-parity violating supersymmetry (SUSY). We calculate
the BRs and the direct CP asymmetries for all the $B \to K \pi$
modes in the SM and its SUSY extension with R-parity (SUGRA models)
and without R-parity. Then, we present predictions of the different
SUSY models to the mixing induced CP violating parameter $S_{K_{_S}
\pi^0}$ which has been observed {\it with large errors} through the
time-dependent CP asymmetry measurement of $B^0 \to K_{_S} \pi^0$
\cite{Aubert:2004dn,Abe:2004xp}. In the recent work
\cite{Kim:2005jp}, it has been explicitly shown that the
color-suppressed tree contribution is very sensitive to the
observable $S_{K_{_S} \pi^0}$, while in contrast, the EW penguin
contribution is not sensitive to $S_{K_{_S} \pi^0}$. As we shall
see, the different SUSY models give different predictions to the
time-dependent CP violating parameter $S_{K_{_S} \pi^0}$ which can
be tested by experiment.

For calculation of the relevant hadronic matrix elements, we adopt the QCD improved factorization
(QCDF) \cite{bbns}. This approach allows us to include the possible non-factorizable contributions,
such as vertex corrections, penguin corrections, hard spectator scattering contributions, and weak
annihilation contributions. The relevant end-point divergent integrals are parameterized as
\cite{bbns}
\begin{eqnarray}
X_{H,A} \equiv \int^1_0 {dx \over x}
 \equiv \left( 1 + \rho_{_{H,A}} e^{i \phi_{_{H,A}}} \right)
 {\rm ln} {m_B \over \Lambda_h} ~,
\label{XAXH}
\end{eqnarray}
where $X_H$ and $X_A$ denote the hard spectator scattering
contribution and the annihilation contribution, respectively.  Here
the phases $\phi_{_{H,A}}$ are arbitrary, $0^0 \leq \phi_{_{H,A}}
\leq 360^0$,  $\rho_{_{H,A}}$ are free parameters to be of order
one, typically $\rho_{_A} \lesssim 2$, and the scale $\Lambda_h =
0.5$ GeV being the typical hadronic scale \cite{bbns}.

We first summarize the current status of the experimental results on $B \to K\pi$
modes in Table~\ref{table:1}, which includes the BRs, the direct CP asymmetries
$({\cal A}_{CP})$, and the mixing-induced CP asymmetry $(S_{K_s \pi^0})$.
In order to exhibit the sign convention for CP asymmetries used in this work, let us
specify the definition of CP asymmetries for $B \to K\pi$ as follows.
The direct CP asymmetry for $B^{\pm} \to K^{\pm} \pi^0$ is defined as
\bea
{\cal A}_{CP}^{+0} &\equiv&
 \frac{{\cal B}(B^- \to K^- \pi^0) -{\cal B}(B^+ \to K^+ \pi^0)}
  {{\cal B}(B^- \to K^- \pi^0) +{\cal B}(B^+ \to K^+ \pi^0)} ~.
\eea
The definition of direct CP asymmetries for other $B \to K\pi$ modes becomes obvious.
The time-dependent CP asymmetry for $B^0 \to K_{_S} \pi^0$ is defined as
\bea
{\cal A}_{K_{_S} \pi^0} (t)
 &\equiv& \frac{\Gamma(\bar B^0 (t) \to K_{_S} \pi^0) -\Gamma(B^0 (t) \to K_{_S} \pi^0)}
  {\Gamma(\bar B^0 (t) \to K_{_S} \pi^0) +\Gamma(B^0 (t) \to K_{_S} \pi^0)}  \nonumber \\
 &\equiv& S_{K_{_S} \pi^0} \sin(\Delta m_d ~t) -C_{K_{_S} \pi^0} \cos(\Delta m_d ~t) ~,
\eea
where $\Gamma$ denotes the relevant decay rate and $\Delta m_d$ is the mass
difference between the two $B^0$ mass eigenstates.
The $S_{K_{_S} \pi^0}$ and $C_{K_{_S} \pi^0}$ are CP violating parameters.
In the case that the tree contributions are neglected for $B^0 \to K_{_S} \pi^0$,
the mixing-induced CP violating parameter $S_{K_{_S} \pi^0}$ is equal to $\sin(2\phi_1)$
[$\phi_1 ~ (\equiv \beta)$ is the angle of the unitarity triangle].
Note that the measured value of $S_{K_{_S} \pi^0}$ (Table~\ref{table:1}) is different from
the well-established value of $\sin(2\phi_1) =0.725 \pm 0.037$ measured through
$B \to J/\psi K^{(*)}$ \cite{HFAG}.
It may indicate that the EW penguin and the color-suppressed tree effects play an important
role \cite{Kim:2005jp}.

\begin{table}
\caption{Experimental data on the CP-averaged branching ratios ($\bar {\cal B}$ in units
of $10^{-6}$), the direct CP asymmetries (${\cal A}_{CP}$), and the mixing-induced CP
asymmetry ($S_{K_s \pi^0}$) for $B \to K\pi$ modes.
The $S_{K_s \pi^0}$ is equal to $\sin(2\phi_1)$ in the case that tree amplitudes are
neglected for $B^0 \to K_s \pi^0$
\cite{HFAG,Bornheim:2003bv,Chao:2003ue,Aubert:2002jb,Aubert:2004dn,Aubert:2004km,
Aubert:2004kn,Abe:LP05,Chen:2000hv,Aubert:2004qm,Chao:2004jy,Chao:2004mn,Abe:2004xp}.}
\smallskip
\begin{tabular}{|c|c||c|c|}
\hline
 BR & Average & CP asymmetry & Average  \\
\hline
$\bar {\cal B}(B^{\pm} \to K^0 \pi^{\pm})$ & $24.1 \pm 1.3$
 & ${\cal A}_{CP}^{0+}$ & $-0.02 \pm 0.04$  \\
$\bar {\cal B}(B^{\pm} \to K^{\pm} \pi^0)$ & $12.1 \pm 0.8$
 & ${\cal A}_{CP}^{+0}$ & $+0.04 \pm 0.04$  \\
$\bar {\cal B}(B^0 \to K^{\pm} \pi^{\mp})$ & $18.9 \pm 0.7$
 & ${\cal A}_{CP}^{+-}$ & $-0.115 \pm 0.018$  \\
$\bar {\cal B}(B^0 \to K^0 \pi^0)$ & $11.5 \pm 1.0$
 & ${\cal A}_{CP}^{00}$ & $+0.001 \pm 0.155$  \\
&  & $S_{K_s \pi^0}$ & $+0.34 \pm 0.29$  \\
\hline
\end{tabular}
\label{table:1}
\end{table}

In the following two sections, we will discuss possible resolutions of the $B \to K\pi$
puzzles in the context of SUSY models.
\\

{\bf [1] R-parity violating SUSY case}

In the R-parity violating (RPV)  minimal supersymmetric standard model, we will assume only
$\l'-$type  couplings to be present \cite{Choudhury:1998wc}.
The R-parity violating interaction introduces new operators.  The relevant new operators
are
\bea
{\cal L}_{eff}
&=& - \frac{\lambda^{\pr}_{i12} \lambda^{\pr *}_{i13}}{2 m^2_{\tilde e_i}}
 \left( \bar u_{\alpha} \gamma_{\mu} L u_{\beta} \right)
 \left( \bar s_{\beta} \gamma_{\mu} R b_{\alpha} \right)
 - \frac{\lambda^{\pr}_{i11 (i32)} \lambda^{\pr *}_{i23 (i11)}}{2 m^2_{\tilde \nu_i}}
 \left( \bar s_{\alpha} \gamma_{\mu} L(R) d_{\beta} \right)
 \left( \bar d_{\beta} \gamma_{\mu} R(L) b_{\alpha} \right)  \no \\
&\mbox{}& - \frac{\lambda^{\pr}_{i12 (i31)} \lambda^{\pr *}_{i13 (i21)}}{2 m^2_{\tilde \nu_i}}
 \left( \bar d_{\alpha} \gamma_{\mu} L(R) d_{\beta} \right)
 \left( \bar s_{\beta} \gamma_{\mu} R(L) b_{\alpha} \right) ~,
\label{Leff}
\eea
where $L(R) = (1 \mp \gamma_5) /2$, $\alpha$ and $\beta$ are the color indices, and $m_{\tilde f}$
denotes the sfermion mass.
Note that the operators having the following chirality structure
$\left( \bar p_{\alpha} \gamma_{\mu} L q_{\beta} \right)
\left( \bar r_{\beta} \gamma_{\mu} R b_{\alpha} \right)$
do not exist in the SM effective Hamiltonian.

The RPV SUSY part of the decay amplitudes of $B \to K \pi$ modes are
given by \cite{Ghosh:2001mr}
\bea
A^{\rm RPV} (\bar B^0 \to K^- \pi^+)
 &=& - i f_K F^{B \to \pi}_0 (0) \left( m_B^2 -m_{\pi}^2 \right) u^R_{112} R_K c_A
  + A^{\rm RPV}_{ann} (K^- \pi^+) ~,
\label{AKmpip}  \\
A^{\rm RPV} (B^- \to K^- \pi^0)
 &=& i f_{\pi} F^{B \to K}_0 (0) \left( m_B^2 -m_K^2 \right)
 \left[ u^R_{112} \frac{1}{\sqrt{2}} \left( -r_{K\pi} R_K c_A + a^{\pr} \right) \right. \no
 &\mbox{}& \left. + \left( d^R_{112} - d^L_{121} \right) \frac{1}{\sqrt{2}} R_{\pi} c_A
 - \left( d^R_{121} - d^L_{112} \right) \frac{1}{\sqrt{2}} a^{\pr}  \right]  \no
 &\mbox{}& + A^{\rm RPV}_{ann} (K^- \pi^0) ~,
\label{AKmpi0} \\
A^{\rm RPV} (B^- \to \bar K^0 \pi^-)
 &=& i f_K F^{B \to \pi}_0 (0) \left( m_B^2 -m_{\pi}^2 \right)
 \left[ \left( d^R_{112} - d^L_{121} \right) a^{\pr}
 - \left( d^R_{121} - d^L_{112} \right) R_K c_A \right]  \no
 &\mbox{}& + A^{\rm RPV}_{ann} (\bar K^0 \pi^-)  ~,
\label{AK0pim} \\
A^{\rm RPV} (\bar B^0 \to \bar K^0 \pi^0)
 &=& i f_{\pi} F^{B \to K}_0 (0) \left( m_B^2 -m_K^2 \right)  \no
 &\mbox{}& \times \left[ u^R_{112} \frac{1}{\sqrt{2}} a^{\pr}
  - \left( d^R_{112} - d^L_{121} \right)
  \frac{1}{\sqrt{2}} ( - R_{\pi} c_A +r_{K\pi} a^{\pr} )  \right.  \no
 &\mbox{}& \left. - \left( d^R_{121} - d^L_{112} \right)
  \frac{1}{\sqrt{2}} ( -r_{K\pi} R_K c_A + a^{\pr})  \right]
  + A^{\rm RPV}_{ann} (\bar K^0 \pi^0) ~,
\label{AK0pi0}
\eea
where the annihilation contributions are given by
\bea
A^{\rm RPV}_{ann} (K^- \pi^+) &=& - \sqrt{2} A^{\rm RPV}_{ann} (\bar K^0 \pi^0)
 = - if_B f_{\pi} f_K \left[ \left( d^R_{112} -d^L_{121} \right) b_4^{\pr}
  + \left( d^R_{121} -d^L_{112} \right) b_3^{\pr} \right] ~,  \\
A^{\rm RPV}_{ann} (\bar K^0 \pi^-) &=& \sqrt{2} A^{\rm RPV}_{ann} (K^- \pi^0)
 = - if_B f_{\pi} f_K ~ u^R_{112} b_3^{\pr} ~.
\eea The $u^R_{jkn}$ and $d^{L,R}_{jkn}$ are defined as $u^R_{jkn} =
\sum_{i=1}^3 {\l'_{ijn}\l'^{\ast}_{ik3} \over 8\msells}$,
$d^R_{jkn}= \sum_{i=1}^3 {\l'_{ijk}\l'^{\ast}_{in3} \over 8\msnus}$,
$d^L_{jkn}= \sum_{i=1}^3 {\l'_{i3k}\l'^{\ast}_{inj} \over 8\msnus}$.
We refer to Refs. \cite{Choudhury:1998wc} for the relevant
notations. Here $f_i$ and $F^{B \to i}_0$ denote decay constants and
form factors, respectively. The parameters $a^{\pr}$, $R_i$, $r_i$
are defined as \bea && a^{\pr} = \frac{c_A}{N_c} \left[ 1 -\frac{C_F
\alpha_s}{4\pi} V_{P_2}^{\pr} \right]
  - \frac{c_A}{N_c} \frac{C_F \pi \alpha_s}{N_c} H_{P_2 P_1}^{\pr} ~, \\
&& R_K = \frac{2 m_K^2}{\bar m_b (\mu) (\bar m_q (\mu) +\bar m_s (\mu))} ~,
  ~~~ (q = u ~(d) ~~ {\rm for} ~~ K^- ~(\bar K^0))  \\
&& R_{\pi} = \frac{2 m_{\pi}^2}{\bar m_b (\mu) (\bar m_u (\mu) +\bar m_d (\mu))} ~, \\
&& r_{K\pi} = \frac{f_K F^{B \to \pi}_0 (0) (m_B^2 -m_{\pi}^2)}
  {f_{\pi} F^{B \to K}_0 (0) (m_B^2 -m_K^2)} ~,
\eea
where $N_c ~(= 3)$ is the number of colors and $C_F = (N_c^2 -1) / (2 N_c)$.
$V_{P_2}^{\pr}$ and $H_{P_1 P_2}^{\pr}$ come from the vertex corrections and the hard
spectator scattering contributions, respectively.
For their explicit expressions, we refer to \cite{beneke}.
$P_1$ is the final state meson absorbing the light spectator quark from B meson and $P_2$
is the other final state meson emitted without absorbing the spectator quark.
The parameters $b_i^{\pr}$ are defined as
\bea
b_3^{\pr} = \frac{C_F}{N_c^2} c_{_C} A_3^f ~, ~~~~~
b_4^{\pr} = \frac{C_F}{N_c^2} c_{_C} A_2^f ~,
\eea
where
\bea
&& A_2^i = \pi \alpha_s \left[ 18 \left( X_A -4 + \frac{\pi^2}{3} \right)
 +2 r_{\chi}^2 X_A^2 \right] ~,  \no
&& A_3^f = 12 \pi \alpha_s r_{\chi} (2 X_A^2 - X_A) ~,
\eea
with $r_{\chi} \approx R_{\pi}$.  $X_A$ is the divergent integral as
defined in Eq. (\ref{XAXH}). $c_{_{A,C}}$ are the RGE improved QCD enhanced factors
at the scale $\mu = m_b$.

From Eqs.~(\ref{AKmpi0})~$-$~(\ref{AK0pi0}), we note that the R-parity
violating couplings $d^R_{ijk}$ and $d^L_{lmn}$ always appear as the
combinations $\left( d^R_{112} -d^L_{121} \right)$ and $\left(
d^R_{121} -d^L_{112} \right)$. Thus, in this analysis, we actually
use three different combinations of R-parity violating couplings:
$u^R_{112}$, $\left( d^R_{112} -d^L_{121} \right)$ and $\left(
d^R_{121} -d^L_{112} \right)$. Since each combination can be
expressed as a complex number, we have six independent real
parameters arising from the new physics effects and we have 9
results to explain. The contributions of the new terms to the
amplitudes are mostly different for different decay modes.

\begin{figure}
\begin{tabular}{cc}
 \DESepsf(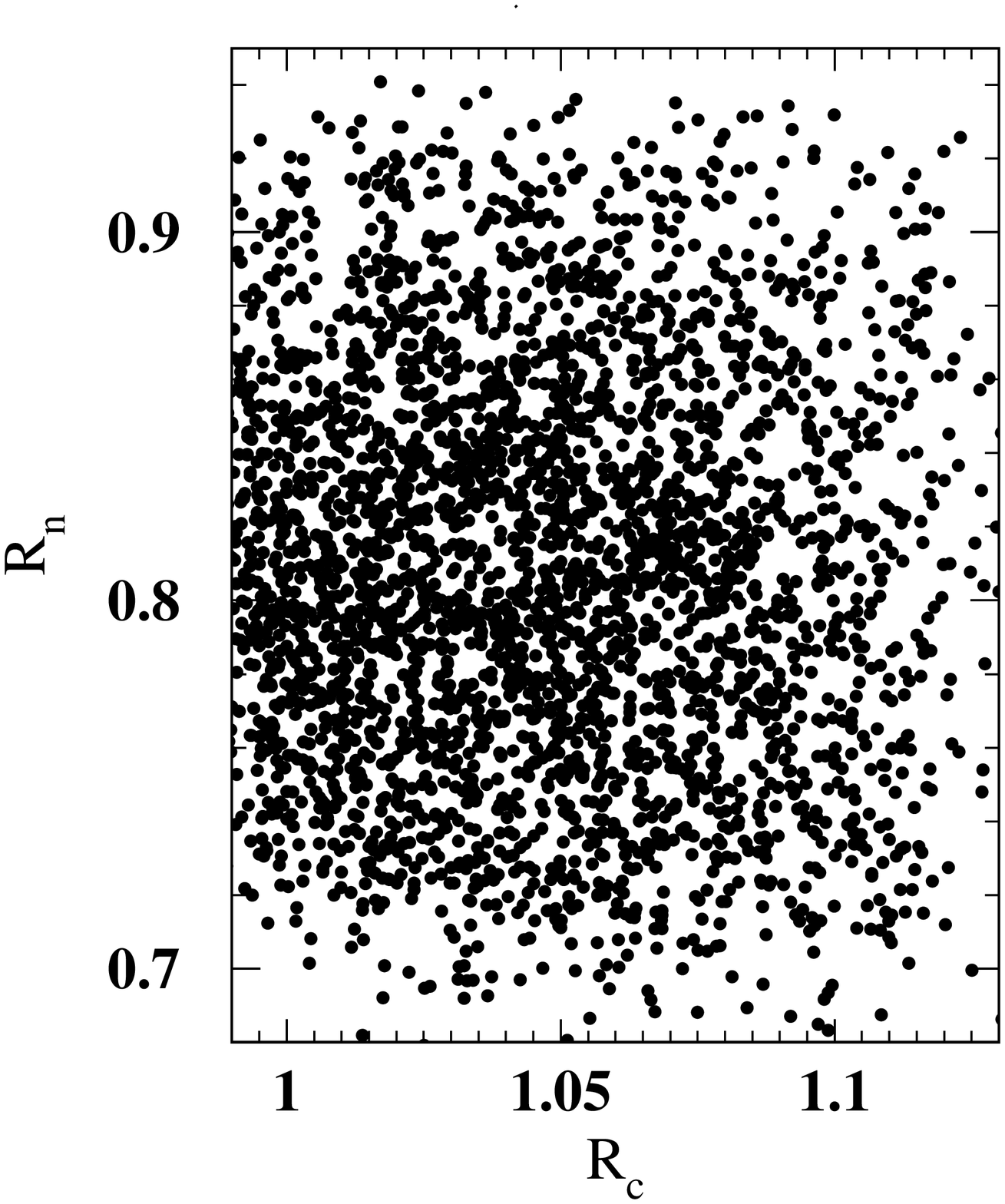 width 5cm) ~~~~~ & ~~~~~
 \DESepsf(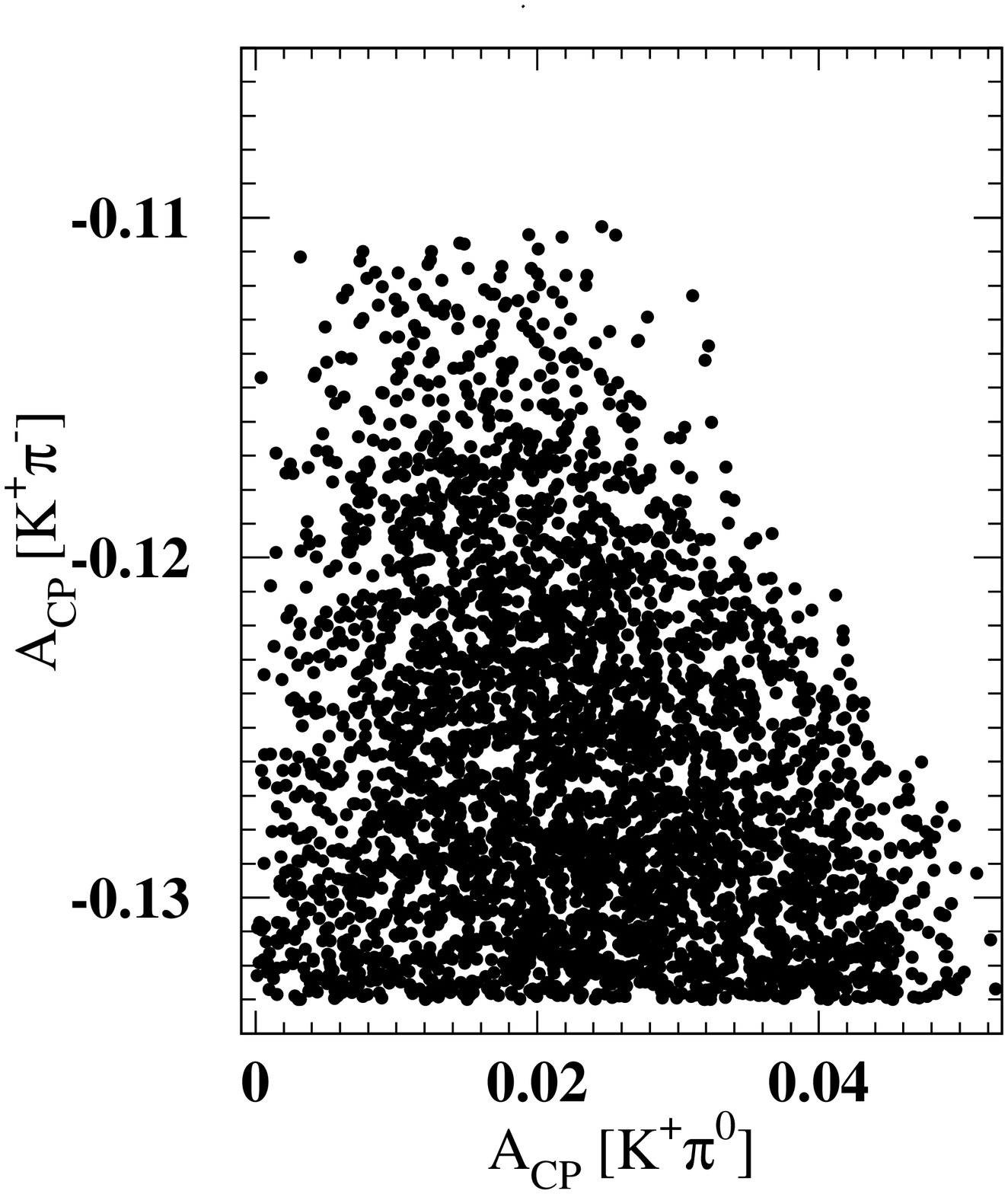 width 5cm)
 \\[-2.0ex]
\textbf{}&\textbf{}
\end{tabular}
\vspace*{8pt}
\caption{ $R_n$ versus $R_c$ (left one) and $A_{CP}^{+-}$ versus $A_{CP}^{+0}$ (right one)
 in the R-parity violating SUSY model.}
\label{fig:1}
\end{figure}

By varying the above parameters, we try to fit all the current data {\it simultaneously}
as shown in Table~\ref{table:1}.
In Fig.~\ref{fig:1}, we show $R_n$ versus $R_c$ (left figure) and $A_{CP}^{+-}$ versus
$A_{CP}^{+0}$ (right figure).  Here the same parameter sets are used to fit both the BRs
and the direct CP asymmetries.  We see that the values of $R_n$, $R_c$, $A_{CP}^{+-}$, and
$A_{CP}^{+0}$ are consistent with the current data at 1$\sigma$ level.
In fact, it turns out that all the current data for the BRs and the direct CP asymmetries,
including $A_{CP}^{0+}$ and $A_{CP}^{00}$, can be explained at 1$\sigma$ level in the
R-parity violating SUSY model.
In other words, the possible discrepancy between the SM predictions and the current data
for the BRs and the direct CP asymmetries can be explained by the new physics contributions
which, in particular, come from the new operators having the new chirality structure as
mentioned below Eq.~(\ref{Leff}).

\begin{figure}
\begin{tabular}{cc}
 \DESepsf(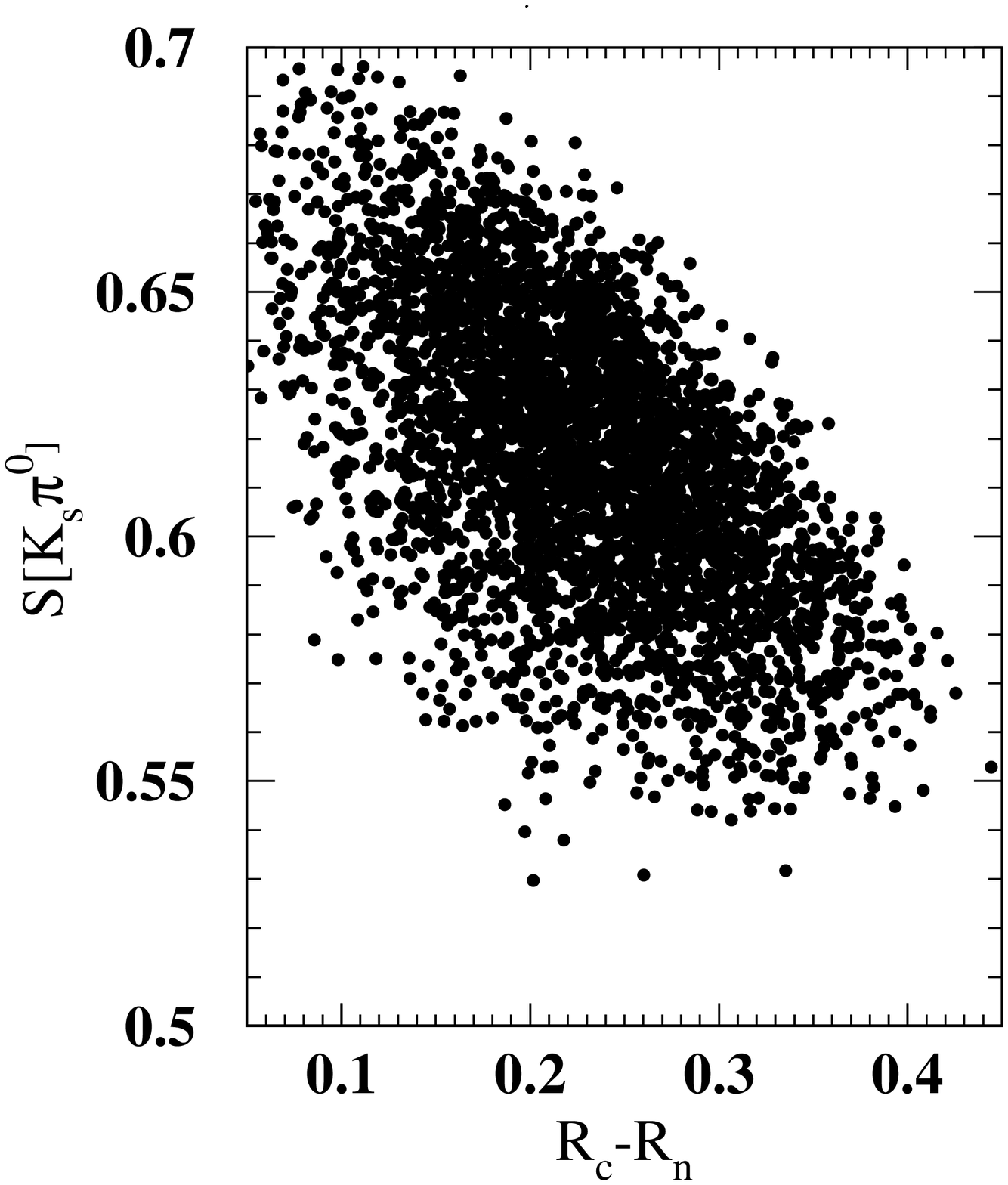 width 5cm) ~~~~~ & ~~~~~
 \DESepsf(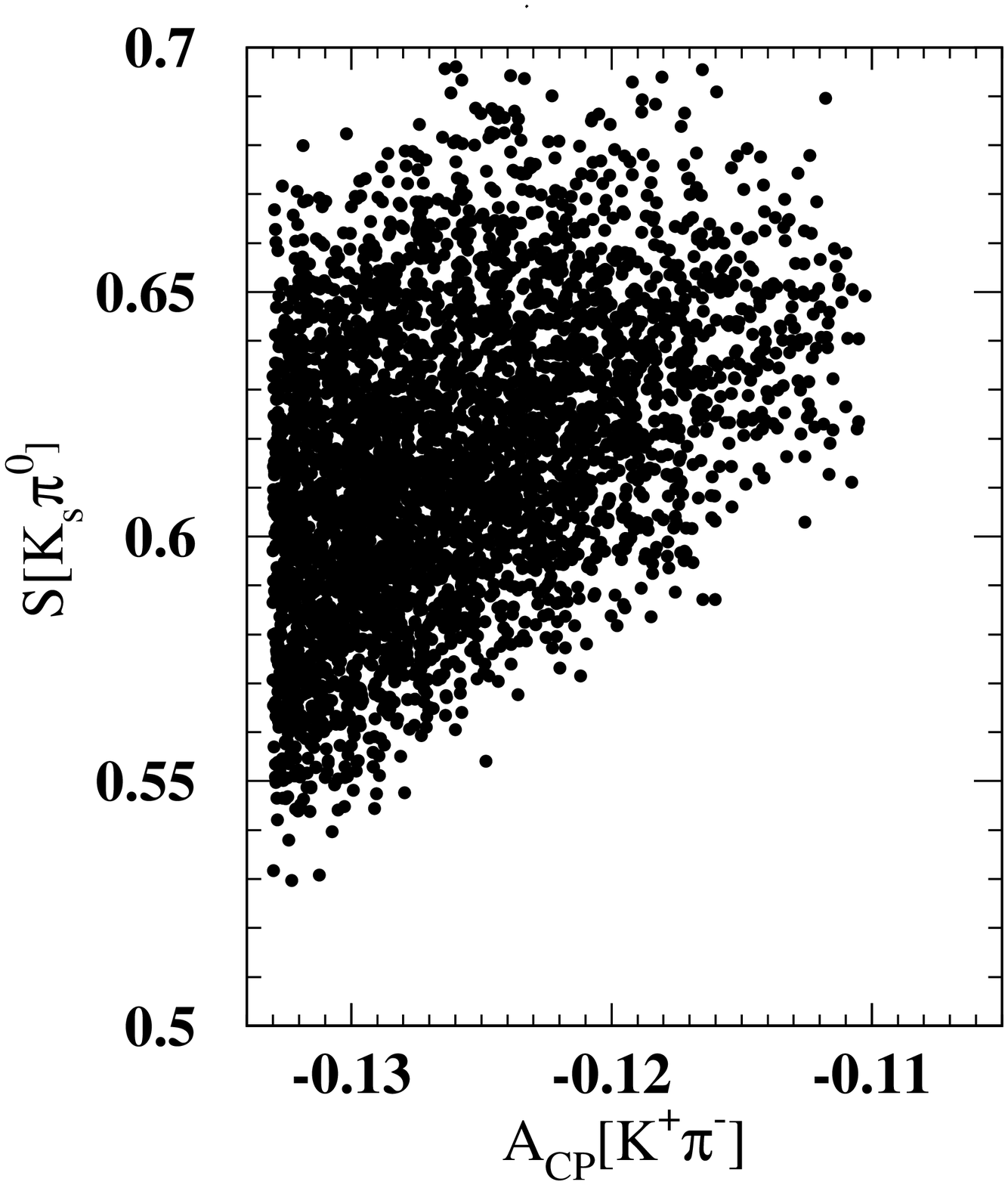 width 5cm)
 \\[-2.0ex]
\textbf{}&\textbf{}
\end{tabular}
\vspace*{8pt}
 \caption{$S_{K_{_S} \pi^0}$ versus $(R_c - R_n)$ (left one) and $S_{K_{_S} \pi^0}$ versus
  $A_{CP}^{+-}$ (right one) in the R-parity violating SUSY model.}
\label{fig:2}
\end{figure}

Using the same values of the parameters used in Fig.~\ref{fig:1}, we predict the mixing
induced CP violating observable $S_{K_{_S} \pi^0}$.  In Fig.~\ref{fig:2}, our result
is presented as $S_{K_{_S} \pi^0}$ versus $(R_c - R_n)$ (left figure) and $S_{K_{_S} \pi^0}$
versus $A_{CP}^{+-}$ (right figure).
We see that our prediction is in good agreement with the current data at 1$\sigma$ level.
We shall see in next section that in R-parity conserving SUSY case, it is very difficult to
explain the small value of the current data for $S_{K_{_S} \pi^0}$ together with the other
data, especially $R_c$ and $R_n$.

In Table~\ref{table:2}, we show the representative values of our
prediction to the BRs, the direct CP asymmetries and the mixing
induced CP asymmetry in the R-parity violating SUSY model. We
consider two cases: (i) $\rho_{_{H,A}} =0$ and (ii) $\rho_{_A} =0.3,
~\rho_{_H} =\phi_{_{H,A}} =0$. The corresponding values of the
couplings are (in $10^{-8}$)
\bea
|d^R_{112} -d^L_{121}| \sim 3.2~(3.1), ~~
|d^R_{121}-d^L_{112}| \sim 0.87~(0.60), ~~
|u^R_{112} | \sim 2.1~(2.2)
\label{couplings}
\eea
The values in the parenthesis are for the $\rho_{_A}=0.3$ case. The
constraints on the RPV couplings need to be checked. However, apart
from $u^R_{112}$, the rest of the couplings appears in the amplitude
as combinations (e.g., $d^R_{112} -d^L_{121}$) of 3 or 4 different
RPV couplings $\lambda^{\pr}_{ijk}$ so that they easily satisfy the
constraints. $u^R_{112}$ involves $\lambda^{\pr}_{i12}\lambda^{\pr
*}_{i13}$. In our example above (for $\rho_{_{H,A}}$=0 case),
$\lambda^{\pr}_{31k} \sim 8 \times 10^{-2}$ was used.  It is also
important to note that  $u^R_{112}$ involves $m_{\tilde e}^2$ which
we assume to be $\sim 200$ GeV.
The experimental bound on $\lambda^{\pr}_{31k}$ is given by
$\lambda^{\pr}_{31k} < 1.2 \times 10^{-1}$ for 1 TeV of squark mass
by using the ratio of BRs of $K^+ \rightarrow \pi^+ \nu \bar \nu$
and $K^+ \rightarrow \pi^0 \nu e^+$ decay \cite{Agashe:1995qm}.
However, the bound on $\lambda^{\pr}$  determined from the
experimental value of the BR of $K \rightarrow \pi \nu \bar \nu$
decay depends on the squark mass and in GUT models, it is quite
natural to expect a large hierarchy ($\sim 5$) between the squark
and the slepton masses.

\begin{table}
\caption{Predictions of the R-parity violating SUSY model for two cases: (i) $\rho_{_{H,A}} =0$
and (ii) $\rho_{_A} =0.3, ~\rho_{_H} =\phi_{{H,A}} =0$.  The case (ii) are shown in the bracket.
($\bar {\cal B}$ in units of $10^{-6}$)}
\smallskip
\begin{tabular}{|c|c||c|c|}
\hline
 BR & Prediction & CP asymmetry & Prediction  \\
\hline
$\bar {\cal B}(B^{\pm} \to K^0 \pi^{\pm})$ & $23.6 ~[24.8]$
 & ${\cal A}_{CP}^{0+}$ & $-0.010 ~[-0.007]$  \\
$\bar {\cal B}(B^{\pm} \to K^{\pm} \pi^0)$ & $13.3 ~[13.0]$
 & ${\cal A}_{CP}^{+0}$ & $+0.026 ~[+0.018]$  \\
$\bar {\cal B}(B^0 \to K^{\pm} \pi^{\mp})$ & $19.0 ~[18.9]$
 & ${\cal A}_{CP}^{+-}$ & $-0.134 ~[-0.115]$  \\
$\bar {\cal B}(B^0 \to K^0 \pi^0)$ & $11.9 ~[12.6]$
 & ${\cal A}_{CP}^{00}$ & $-0.142 ~[-0.141]$  \\
&  & $S_{K_s \pi^0}$ & $+0.51 ~[+0.55]$  \\
\hline
\end{tabular}
\label{table:2}
\end{table}

{\bf [2] R-parity conserving SUSY case}

In this case the SUSY contributions appear in loop. The one loop
SUSY contributions are available in the literature, e.g., Refs.
\cite{Bertolini:1990if,Gabbiani:1996hi}. In our calculation, we do
not use the mass insertion approximation, but rather do a complete
calculation. The SUGRA model starts at the GUT scale. We assume the
breakdown of the universality to accommodate the $B\rightarrow \pi
K$ data.  While we satisfy this data, we also have to be careful to
also satisfy other data, e.g., $b\rightarrow s\gamma$, $\Delta M_K$,
$\Delta B_d$, $\epsilon_K$, etc.

We use the following boundary conditions at the GUT scale:
\begin{equation}
( m^{2} )^{ij}_{(Q_{LL},U_{RR},D_{RR})}
 = m_0^2 \left( \delta^{ij} +\Delta^{ij}_{(Q_{LL},U_{RR},D_{RR})} \right) ~; ~~~
A^{ij}_{(u,d)} = A_0 \left( Y^{ij}_{(u,d)} +\Delta A^{ij}_{(u,d)} \right) ~.
\end{equation}
The SUSY parameters can have phases at the GUT scale:
$m_{i}=|m_{1/2}| e^{i\theta_i}$ ($i=1,2,3$) (the gaugino masses for
the $U(1)$, $SU(2)$ and $SU(3)$ groups), $A_0=|A_0|e^{i\alpha_{_A}}$
and $\mu=|\mu| e^{i\theta_\mu}$. However, we can set one of the
gaugino phases to zero and we choose $\theta_2 =0$. The electric
dipole moments (EDMs) of the electron and neutron can now allow the
existence of large phases in the theory \cite{edm1,edm2,AAB1}.  In
our calculation, we use $O(1)$ phases but calculate the EDMs to make
sure that current bounds ($|d_e|<1.2\times
10^{-27}$ecm~\cite{Regan:2002ta} and $|d_n|<6.3\times
10^{-26}$ecm~\cite{Harris:1999jx}) are satisfied.

We evaluate the squark masses and mixings at the weak scale by using
the above boundary conditions at the GUT scale. The RGE evolution
mixes the non-universality of type LR (A terms) via $d
{m_Q}^2_{LL,RR} /dt \propto A_{u(d)}^{\dagger}A_{u(d)}$ terms and
creates new LL and RR contributions at the weak scale.  We then
evaluate the Wilson coefficients from all these new contributions.
We have both chargino and gluino contributions arising due to the
LL, LR, RL, RR up type and down type squark mixing.  These
contributions affect the following Wilson coffecients $C_3-C_{10}$,
$C_{7\gamma}$ and $C_{8g}$.  The chargino contributions affect
mostly the electroweak penguins ($C_7$ and $C_9$) and the dipole
penguins, while the gluino penguin has a large contribution to the
dipole terms due to the presence of an enhancement factor $m_{\tilde
g}/m_b$ (the gluino contribution also affects the QCD penguins, but
the effect is small). We include all contributions in our
calculation. The SUSY contributions also bring new operator
contributions over the SM by having a chirality exchange in the SM
operators.

\begin{table}
\caption{Predictions of the R-parity conserving SUSY model. The SUSY parameters
are mentioned in the text. ($\bar {\cal B}$ in units of $10^{-6}$)}
\smallskip
\begin{tabular}{|c|c||c|c|}
\hline
 BR & Prediction & CP asymmetry & Prediction  \\
\hline $\bar {\cal B}(B^{\pm} \to K^0 \pi^{\pm})$ & $23 $
 & ${\cal A}_{CP}^{0+}$ & $-0.030 $\\
$\bar {\cal B}(B^{\pm} \to K^{\pm} \pi^0)$ & $10.3 $
 & ${\cal A}_{CP}^{+0}$ & $-0.0073$\\
$\bar {\cal B}(B^0 \to K^{\pm} \pi^{\mp})$ & $19.1 $
 & ${\cal A}_{CP}^{+-}$ & $-0.105$  \\
$\bar {\cal B}(B^0 \to K^0 \pi^0)$ & $11.3$
 & ${\cal A}_{CP}^{00}$ & $-0.08 $\\
&  & $S_{K_s \pi^0}$ & $+0.73$\\
\hline
\end{tabular}
\label{table:3}
\end{table}

The electroweak penguin contribution is required to solve the
$B\rightarrow \pi K$ puzzle for the BRs and can solve the CP asymmetries
\cite{Buras:2004th}. If we do not consider the BRs, then the direct
CP asymmetries of the $B\rightarrow \pi K$ modes can be solved by
the dipole penguin contributions only. The dipole penguin
contributions can not be arbitrarily large, since it is also present
in the $b\rightarrow s\gamma$.  In order to obtain a fit, we find
that $A^{23}_{u,d}$ are necessary. The nonzero values of these
parameters generate the dipole penguin and the ($Z$-mediated)
electroweak penguin  diagrams. In Table \ref{table:3}, we show an
example of a fit. From the fit one finds the prediction for $S_{K_s
\pi^0}$ to be large. The SUSY parameters used for this fit are:
$m_{1/2} =450$ GeV, $A_0 =-800$ GeV, $m_0 =300$ GeV,
$\Delta_{Q_{LL}}^{23} =0.2 ~e^{-0.3 i}$, $\Delta A_{u}^{23} =0.55
~e^{0.8 i}$, $\Delta A_{d}^{23} =0.05 ~e^{-1.5 i}$, $\tan \beta
=40$, $\mu >0$. Since the SUSY parameters have phases, the EDMs of
the electron and the neutron need to be checked, and we do indeed
satisfy the experimental bounds for these EDMs. For this example, we
find $|d_e| =2.23 \times 10^{-29}$ e cm and $|d_n| =8.2 \times 10^{-27}$ e cm.
The QCD parameters for this fit are: $\rho_{_A} =2$
and $\phi_{_A} =2.77$. In this fit we have used nonzero
$\Delta_{Q_{LL}}$, but it is possible to obtain fits without
$\Delta_{Q_{LL}}$. We can obtain fits for other $\tan\beta$ values
as well.

\begin{figure}
\begin{tabular}{cc}
 \DESepsf(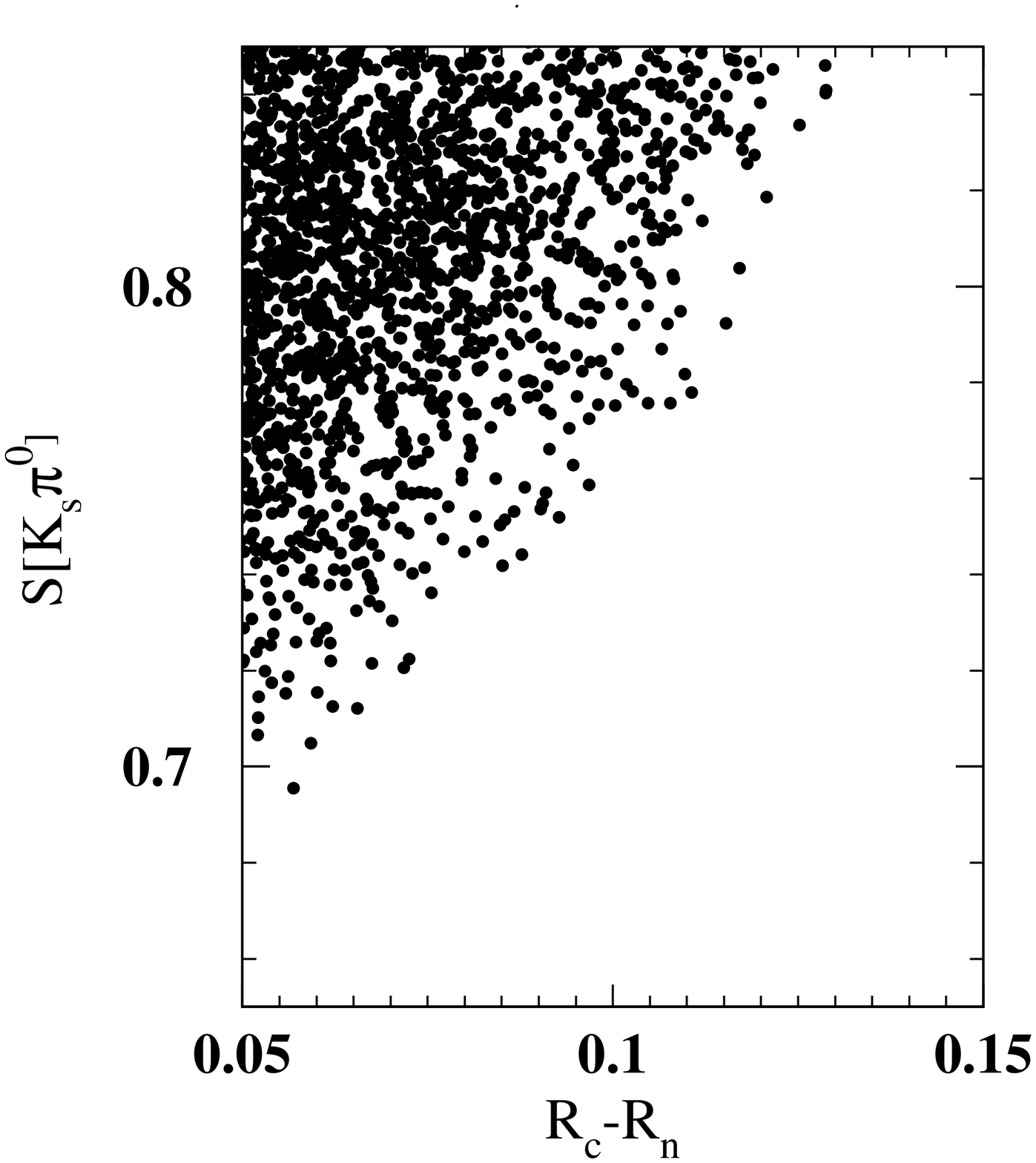 width 5cm) ~~~~~ & ~~~~~
 \DESepsf(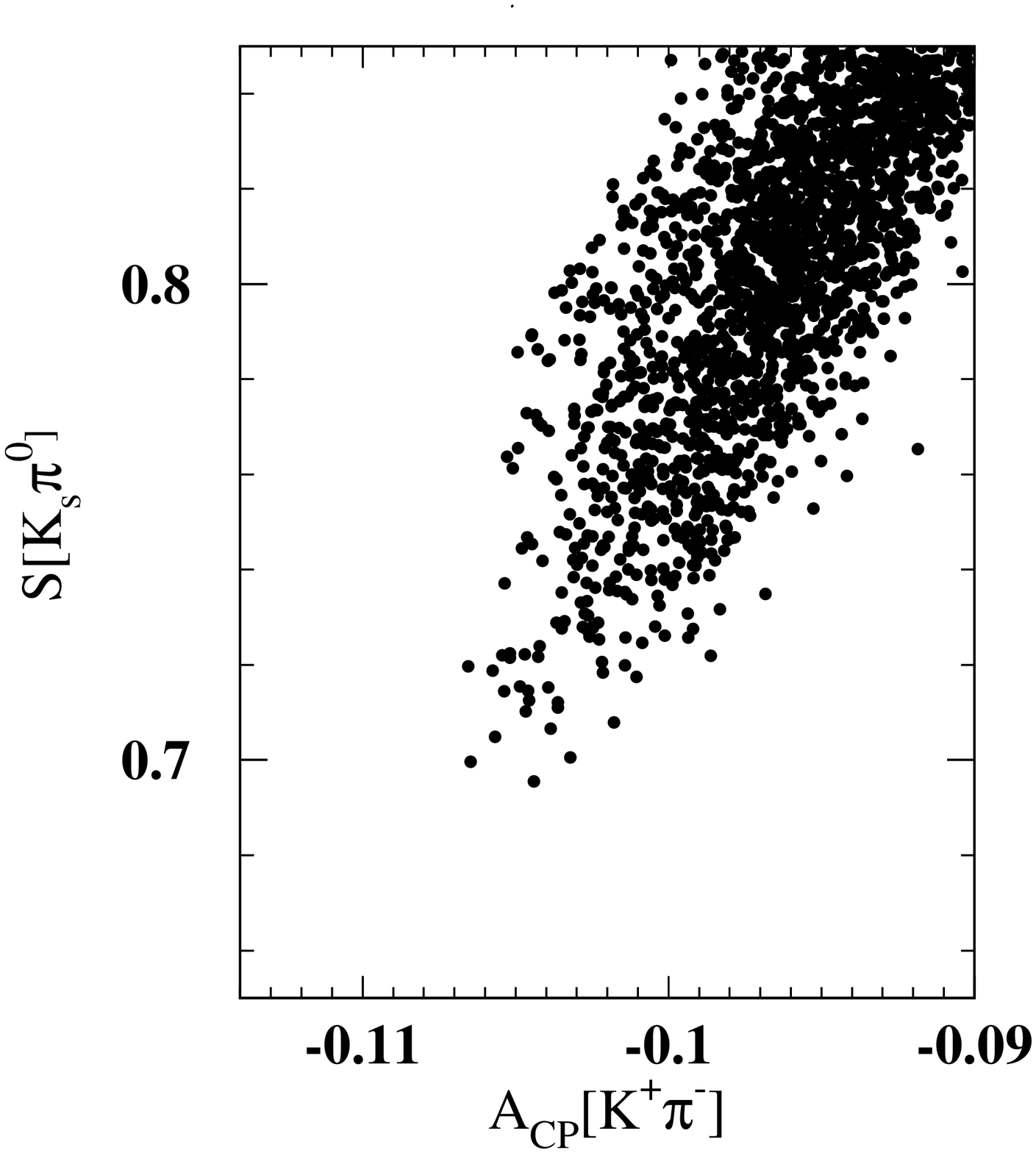 width 5cm)
 \\[-2.0ex]
\textbf{}&\textbf{}
\end{tabular}
\vspace*{8pt}
 \caption{$S_{K_{_S} \pi^0}$ versus $(R_c - R_n)$ (left one) and $S_{K_{_S} \pi^0}$
  versus $A_{CP}^{+-}$ (right one) in the SUGRA model.}
\label{fig:3}
\end{figure}

In Fig.3, we show $S_{K_s \pi^0}$ as a function of $(R_c -R_n)$
and $S_{K_s \pi^0}$ as a function of $A_{CP}^{+-}$. In order to
generate these figures, we have varied $m_{1/2}$, $m_0$,
$\tan\beta$ and $\Delta$'s. We see from the figures that the
lowest value of $S_{K_s \pi^0}$ is about 0.69 and the maximum direct CP
asymmetry $A_{CP}^{+-}$ predicted by the SUGRA model is about $-0.107$.
If we compare Figure 3 with Figure 2, we find that the prediction for
$S_{K_s \pi^0}$ in the R-parity conserving SUSY model is much
higher than in the R-parity violating SUSY model and therefore the
future data on $S_{K_s \pi^0}$ will be crucial. The future data
(with reduced error) of $A_{CP}^{+-}$ is also crucial to
distinguish two scenarios since the maximum direct CP asymmetry $A_{CP}^{+-}$
predicted by the SUGRA model is about $-0.107$, whereas the asymmetry can be
larger negative in the R-parity violating model.

In conclusion, we have explained the recent experimental results on
the BRs and CP asymmetries of different $B\rightarrow \pi K$ modes
in R-parity violating and R-parity conserving SUSY models. We have
found that the R-parity conserving SUSY model tends to generate
large $S_{K_{_S} \pi^0}$ when we use all the constraints on the BRs
and CP asymmetries, and the lowest value of $S_{K_{_S} \pi^0}$ is
about 0.69. However,  lower values of $S_{K_{_S} \pi^0}$ can be
accommodated in the R-parity violating SUSY model. We also find that
the future data of $A^{+-}_{CP}$ is important to distinguish the two
models.

\vspace{1cm} \centerline{\bf ACKNOWLEDGEMENTS} \noindent The work of
S.O. was supported by the Korea Research Foundation Grant
(KRF-2004-050-C00005)


\end{document}